\documentclass[11pt]{article}

\usepackage{epsfig}
\usepackage{latexsym}
\usepackage{amssymb}

\newcommand{\sect}[1]{\setcounter{equation}{0}\section{#1}}

\newcommand{\reef}[1]{(\ref{#1})}
\def\be{\begin{equation}}
\def\ee{\end{equation}}
\def\ba{\begin{eqnarray}}
\def\ea{\end{eqnarray}}
\def\prt{\partial}

\begin{document}

\thispagestyle{empty}
\rightline{\small hep-th/0206243 \hfill MIT-CTP-3287}
\rightline{\small \hfill  RUNHETC-2002-22} 
\vspace*{2cm}

\begin{center}
{ \LARGE {\bf Calibrations, Monopoles and Fuzzy Funnels  }}\\[.25em]
\vspace*{1cm}

Neil R. Constable$^a$\footnote{E-mail: constabl@lns.mit.edu}
and Neil D. Lambert$^b$\footnote{E-mail: nlambert@physics.rutgers.edu}
\\
\vspace*{0.2cm}
${}^a${\it Center for Theoretical Physics and Laboratory for Nuclear Science}\\
{\it Massachusetts Institute of Technology}\\
{\it Cambridge, Ma. 02139} \\
{\it USA}\\
\vspace*{0.2cm}
${}^b${\it Department of Physics and Astronomy}\\
{\it Rutgers University}\\  
{\it Piscataway NJ 08855}\\ 
{\it USA}\\

\vspace{2cm} ABSTRACT
\end{center}
\noindent
We present new non-Abelian solitonic configurations in the low energy effective
theory describing a collection of $N$ parallel D1--branes. These configurations
preserve $1/4$, $1/8$, $1/16$ and $1/32$ of the spacetime supersymmetry.
They are solutions to a set of generalised Nahm's equations 
which are related to 
self-duality equations in eight dimensions. Our solutions represent D1--branes
which expand into fuzzy funnel configurations ending on collections of 
intersecting D3--branes. Supersymmetry dictates that such 
intersecting D3--branes must lie on a calibrated three-surface of 
spacetime and we argue that the 
generalised Nahm's equations encode the data for the construction of 
magnetic monopoles on the relevant three-surfaces.

\vfill \setcounter{page}{0} \setcounter{footnote}{0}
\newpage

\sect{Introduction}

In the absence of gauge fields branes are embedded into spacetime so
as to minimize their worldvolume. In addition, requiring that the branes
are supersymmetric leads to the condition that their
worldvolumes are in a preferred class of sub-manifolds of spacetime known as 
calibrated surfaces~\cite{HL}. 
These surfaces were first applied to supersymmetric
brane configurations within string theory compactifications
in ~\cite{BBS,BBM}.
The theory of calibrations is also important for understanding 
intersecting D--branes which preserve some fraction of supersymmetry 
in flat space~\cite{GP,GLW,AFS}. 
In this context the combined worldvolume of a set of intersecting
D$p$--branes can be regarded as a single $p+1$ dimensional hypersurface 
embedded into spacetime. The condition that the embedding is supersymmetric
is equivalent to the condition that the $p+1$ dimensional hypersurface
is calibrated but now these surfaces are generically non-compact. 

In particular it was shown in~\cite{GLW} that
the calibration equations of ~\cite{HL} are realised as a BPS condition
in the low energy effective field theory of a {\it single} 
D$p$--brane. For this
reason we refer to these as {\it Abelian} 
embeddings. As is well known, multiple
coincident D$p$--branes give rise to a non-Abelian gauge symmetry in the low
energy effective theory~\cite{witten}. This has lead to many 
interesting implications for  the spacetime interpretation of D--branes
and has given invaluable new geometrical insights into non-Abelian 
gauge theory phenomena such as monopoles and instantons. 
It is therefore of interest to understand {\it non-Abelian} embeddings of 
D--branes into spacetime. 
Ideally one would like to derive the non-Abelian
generalisation of the BPS conditions obtained for a single D$p$--brane 
in~\cite{GLW}. However this issue is at present complicated 
by the fact that the non-Abelian generalisation of the Born-Infeld 
action is poorly understood and in particular there is virtually 
no understanding of
non-Abelian $\kappa$-symmetry--see \cite{berg1,berg2,howe} for some 
recent discussion and progress on this issue.

In this paper we will initiate a study of non-Abelian embeddings by studying
the simplest example, namely D1--branes. 
Due to the absence of a suitable notion 
of non-Abelian $\kappa$-supersymmetry we will restrict our analysis to
the Yang-Mills approximation. We expect that any 
supersymmetric solution to the Yang-Mills equations of motion can be
lifted to a the 
full non-linear equations of motion. In the study of Abelian embeddings
this corresponds to considering free scalar Maxwell theories and is trivial,
{\it i.e.} most of  the interesting solutions crucially 
involve the non-linear terms.
However we will find that non-trivial solutions exist in the non-Abelian
Yang-Mills approximation.

More specifically 
we will see that the resulting solitons are generalisations of the
D1$\perp$D3 system in which $N$ 
D1--branes end on a single D3--brane.   
From the point of view of the effective theory living on the D3--brane
the D1--branes act as point sources of magnetic charge. These 
monopole configurations are realised as BPS solutions to the equations of 
motion which represent the worldvolume of the D3--brane protruding into
the ambient spacetime in a spike-like configuration known as a 
BI-on~\cite{cm,hlw,gibb}. As a BPS configuration in a $U(1)$ field theory the
BI-on is a completely Abelian object. 
This system can instead be studied as a solitonic
object in the non-Abelian theory describing the $N$ D1--branes.
The BPS equations for this system are known to be given by the
Nahm equations~\cite{nahm} describing BPS monopoles in $3+1$ dimensional
Yang-Mills-Higgs theory~\cite{ded}. The Nahm data used in the construction 
of monopoles in the D3--brane theory is therefore 
naturally encoded in the non-Abelian dynamics of the D1--branes
which end on the D3--branes.   

It was shown in\cite{core} that such solitonic solutions 
in the D1--brane theory in fact 
describe a non-commutative (or fuzzy) funnel configuration which opens up into
a D3--brane orthogonal to the worldvolume of the D1--branes. Further, it was 
also shown that this funnel actually acts as a source for the Ramond-Ramond
four form of type IIB string theory in precisely the correct fashion to be 
identified with the D3--brane on which the D1--branes are ending. This 
construction therefore provides a self-contained description
of the D1$\perp$D3 intersection which is complementary 
to that of the BI-on spike. 

The fuzzy funnel configurations described in \cite{core} are solutions
to Nahm's equations and describe D1--branes ending on a single D3--brane. 
As such, these constructions involve only three of the eight transverse
scalars in the effective theory describing the D1--branes. It is natural
to ask whether one may find analogous solitonic configurations which involve
more, perhaps all, of the transverse scalars. This question was addressed in
\cite{fun} where such solutions involving five scalars were found. These
were shown to correspond to D1--branes ending on a collection of D5--branes. 
While stable these configurations are not supersymmetric. 
It is the purpose of the present investigation to discuss new solitonic
non-Abelian D1--brane configurations which involve five, six 
and seven of the transverse scalars while preserving 
various fractions of the supersymmetry preserved by the original D1--branes. 

In the following we will denote the transverse scalars of the non-Abelian
D1--brane theory by $\Phi^i,\, i=1,\cdots ,8$ and take the configuration
to lie along the $x^9$ direction in spacetime. We will show that there exist
a class of supersymmetric configurations which are solutions to the 
generalised Nahm's equations
\be
\frac{\prt\Phi^i}{\prt x^9}= \frac{1}{2}c_{ijk}[\Phi^j,\Phi^k]\ ,
\label{intronahm}
\ee
where $c_{ijk}$ is a totally antisymmetric constant tensor which we will
determine exactly below. These equations can also be derived as the
dimensional reduction of the higher-dimensional self-duality conditions in
\cite{Corrigan}. In the special case that only three of the transverse
scalars are non-trivial $c_{ijk}$ becomes $\varepsilon_{ijk}$ and 
equation~\reef{intronahm} is simply the standard Nahm's equations discussed in 
~\cite{ded,core} corresponding to the D1--branes ending on a single 
D3--brane. The more general configurations involving more than
three non-Abelian scalars will turn out to correspond to fuzzy funnel 
configurations in which the D1--branes end on collections of intersecting 
D3--branes. As discussed above such intersecting  D3--branes are 
supersymmetric if their worldvolumes lie on a calibrated three-manifold
\cite{HL}. Defining the three form
$\omega = \frac{1}{3!}c_{ijk}dx^i\wedge dx^j\wedge dx^k$ we will show 
that the non-trivial 
components of $\omega$ are in a one to one correspondence with the 
three-form which 
calibrates the D3--brane intersection on which the D1--branes are ending. We
are thus led to conjecture that the generalised Nahm's equations 
encode the necessary data
to construct magnetic monopole solutions on various calibrated three-manifolds
via an analogous method to the standard Nahm construction of magnetic monopoles
on $\mathbb{R}^3$. In addition we will argue that these generalised Nahm 
equations also encode data on the moduli of the associated calibrated
manifold. 

The remainder of this paper is organized as follows. In section two we derive
the generalised Nahm's equations as a BPS condition in the non-Abelian 
worldvolume theory describing $N$ D1--branes. In section three 
we discuss how solutions to the generalized Nahm's equations are related to 
supersymmetric D3--brane configurations. Specifically, we make precise 
our claims that the generalised Nahm's equations encode the data describing
magnetic monopoles on calibrated three-manifolds. We also
discuss the connection between our construction and the notion of 
higher dimensional self dual gauge fields~\cite{Corrigan}. 
In section four we provide explicit solutions to the
generalised Nahm's equations and demonstrate that these solutions can be 
interpreted as fuzzy funnels which open up into the various D3--brane 
configurations discussed in sections two and three. In section five
we examine the moduli associated with our solutions and identify the 
deformations which correspond to the moduli of the calibrated surfaces on 
which the D1--branes are ending. We conclude with some comments and open
problems.

\sect{Generalised Nahm Equations from D1--branes}

The effective action for $N$ D1--branes obtained by quantization of
open strings is a  non-linear generalisation of the
Yang-Mills action with gauge group $U(N)$. We will choose
conventions where the fields, which take values in the Lie Algebra $u(N)$,
are anti-Hermitian.
The worldvolume gauge field has explicitly been set to zero. In the following
we will work in static gauge so that the worldvolume coordinates are 
identified with those of spacetime as, $\sigma^1=t$ and $\sigma^2=x^9$. 
Furthermore we will be considering only the leading order terms appearing in an
expansion of the full non-Abelian Dirac-Born-Infeld effective action
\footnote{We assume here the conventions of ~\cite{dielec}} 
\be\label{loworder}
S= -T_1\int dtdx^9\left(N+\lambda^2{\rm Tr}\left(\frac{1}{2}\prt^a\Phi^i\prt_a\Phi^i
+\frac{1}{4}\left[\Phi^i,\Phi^j\right]\left[\Phi^i,\Phi^j\right] + 
\cdots\right)\right)\ ,
\ee
where $\lambda=2\pi l_s^2$ and $i,j=1,...,D$ 
labels the number of non-vanishing transverse directions.
The terms involving the scalar fields are easily recognised as
the Bosonic sector of maximally supersymmetric
Yang-Mills theory  with gauge group $U(N)$ in two spacetime dimensions
described by the Lagrangian
\be
\label{L}
{\cal L} = T_1\lambda^2{\rm Tr}\left( {1\over 2}\partial_\mu \Phi^i 
\partial^\mu \Phi^i 
+{1\over 4}[\Phi^i,\Phi^j]^2\right)\ .
\ee
After subtracting off the ground state energy of the D1--branes, 
the energy of a static configuration can be written as
\ba
\label{E}
E &=& T_1\lambda^2\int dx^9{\rm Tr}\left( {1\over 2} {\Phi^i}'{\Phi^i}'+
{1\over 4}[\Phi^i,\Phi^j][\Phi^i,\Phi^j]
\right)\nonumber\\
&=&T_1\lambda^2\int dx^9\left\{{1\over 2}{\rm Tr}\left({\Phi^i}' -{1\over 2}
c_{ijk}[\Phi^j,\Phi^k]\right)^2 
+ T\right\}\ , \nonumber\\
\ea
where a
prime denotes differentiation with respect to $x^9$ and 
we have introduced a constant, totally anti-symmetric  
tensor $c_{ijk}$ which will be specified below.
In \reef{E} we have performed the usual Bogomoln'yi construction and
written the energy density as a squared term plus a topological piece
given by
\be
\label{T}
T = {T_1\lambda^2\over 3}c_{ijk}{\rm Tr}\left(\Phi^i\Phi^j\Phi^k\right)'\ .
\ee 
We must also impose that the two quartic terms in \reef{E} agree, that is 
we must impose that 
\be
\label{constraint1}
{1\over 2}c_{ijk}c_{ilm}{\rm Tr}\left([\Phi^j,\Phi^k][\Phi^l,\Phi^m]\right) 
= {\rm Tr}\left([\Phi^i,\Phi^j][\Phi^i,\Phi^j]\right) \ .
\ee 
It now follows that the Bogomoln'yi equation is
\be\label{Beq}
{\Phi^i}' ={1\over 2}c_{ijk} [\Phi^j,\Phi^k]\ ,
\ee
and hence the energy of such a configuration depends only on the
boundary conditions of the coordinates.

Equation \reef{Beq} can be thought of as a generalised Nahm equation.
Indeed, in the case that $\Phi^i\neq 0$ for $i=1,2,3$ and $c_{ijk}=
\varepsilon_{ijk}$
\reef{Beq} is precisely Nahm's equation~\cite{nahm} describing  
BPS monopoles on $\mathbb{R}^3$. The emergence 
of this equation in the low energy D1--brane theory was first pointed out in 
~\cite{ded}. This was further studied in \cite{core} 
where it was found 
that this equation has supersymmetric solutions which represent the 
expansion of the parallel D1--branes into a non-commutative funnel structure
which opens up into an orthogonal D3-brane filling the $x^1,x^2,x^3$ 
directions in spacetime.
As will be 
seen below~\reef{Beq} possesses supersymmetric non-commutative funnel 
solutions, with the appropriate choices of $c_{ijk}$, involving 
$5,6 \,\,{\rm and}\,\, 7$ transverse scalars.
These solutions preserve $1/4,1/8  
\,\,{\rm and }\,\, 1/16$ of the supersymmetry 
preserved by the original collection
of D1--branes. Further, instead of expanding into a single D3-brane, 
these configurations are found to be expanding into collections
of intersecting D3-branes.

In order to proceed note that since \reef{constraint1} is  
a constraint on the fields we must also ensure
that the equation of motion is satisfied
\be\label{eqofm}
{\Phi^i}'' = -[[\Phi^i,\Phi^j],\Phi^j]\ ,
\ee
or, from \reef{Beq},
\be\label{constraint2}
{1\over 2}c_{ijk}c_{jlm}[[\Phi^l,\Phi^m],\Phi^k] =
-[[\Phi^i,\Phi^j],\Phi^j]
\ .
\ee 
Multiplying \reef{constraint2} by $\Phi^i$ and taking the trace 
we find that the
constraint \reef{constraint1} is automatically satisfied. Thus solutions to
the generalized Nahm's equations, along with the constraint \reef{constraint2}, 
are guaranteed to be solutions of the full equations of motion.

We are interested in supersymmetric solutions of equation~\reef{Beq}. 
The general supersymmetry variation is
\be\label{gensusy}
\delta\lambda = ({1\over 2}\partial_\mu \Phi_i\Gamma^{\mu i}
    +{1\over 4}[\Phi^i,\Phi^j]\Gamma^{ij})\epsilon\ .
\ee
Here the $\Gamma$-matrices form  the $Spin(1,9)$ Clifford algebra.
In our case $\delta\lambda=0$  becomes
\be
\label{susy}
0=\sum_{i<j}[\Phi^i,\Phi^j]\Gamma^{ij}(1 + c_{ijk}\Gamma^{ijk9})\epsilon\ .
\ee
Note that $\epsilon$ is the preserved supersymmetry on the 
D1--brane worldvolume.
To solve the supersymmetry condition \reef{susy}  we define the projectors
\be
\label{P}
P_{ij}= {1\over 2}(1 + c_{ijk}\Gamma^{ijk9})\ ,
\ee 
where there is no sum on $i,j$. In the cases we will 
consider,  for a given pair $i,j$,  $c_{ijk}$ is only
non-zero for at most one value of $k$. In this
case, provided that we normalize $c_{ijk}=\pm1$, we find that 
$P_{ij}^2=P_{ij}$. Hence we set   
$P_{ij}\epsilon=0$ for each pair $i,j$ such that $c_{ijk}\ne 0$ for some $k$.\footnote{Note that there may be more
general solutions to the supersymmetry condition but we expect that
these are related to those discussed here by a rotation.}
Of course to find a non-trivial solution for $\epsilon$, and hence
preserve some fraction of the D1--brane's sixteen supersymmetries, 
we must impose that the matrices $\Gamma^{ijk9}$ which appear in the $P_{ij}$
projectors commute with each other. It is not hard to see that 
$[\Gamma^{ijk9},\Gamma^{i'j'k'9}]=0$ 
if and only if  $i=i'$ and $j\ne k\ne j'\ne k'$, etc., {\it i.e.} 
the  sets
$\{i,j,k\}$ and $\{i',j',k'\}$ have exactly one element in common.
Note that it is possible that some combination of projectors imply that other
projectors are automatically satisfied.
The number of preserved spacetime 
supersymmetries is $16 \times 2^{-k}$ where $k$ is
the number of independent projectors $P_{ij}$ that we impose.

Once we have a particular set of mutually commuting projectors 
$P_{ij}$ we can return to the supersymmetry transformation \reef{susy}.
This now becomes 
\be
\label{condition}
\sum_{c_{ijk}=0}[\Phi^i,\Phi^j]\Gamma^{ij}\epsilon=0\ ,
\ee 
where the sum is over pairs $i,j$ such that $c_{ijk}= 0$ for all $k$. 
The projectors can then be used to reduce this equation to a set of conditions
on the commutators $[\Phi^i,\Phi^j]$ alone. 
As a consequence of the specific form for the $c_{ijk}$ elements that
we will use, one can show  that \reef{condition} and the supersymmetry
projectors \reef{P} imply  \reef{constraint2}.
To summarise then we must solve the Bogomoln'yi equation \reef{Beq} and 
the commutator conditions that follow from \reef{condition}. 

\sect{D3--branes and Calibrated Geometry}

In this section we wish to understand the spacetime origin of the projectors
obtained in equation \reef{P}.
Recall that in type IIB string theory 
there are two  ten-dimensional supersymmetry
generators $\epsilon_L$ and $\epsilon_R$ with the same
chirality. The presence of the D1--brane breaks half of the supersymmetry of
the vacuum by imposing that $\Gamma^{09}\epsilon_L=\epsilon_R$.
Using this relation the projection
$\Gamma^{ijk9}\epsilon_L = \mp \epsilon_L$ 
can be written as $\Gamma^{0ijk}\epsilon_L=\pm\epsilon_R$. We immediately
recognise this projector as due to the presence of a (anti-)D3--brane
in the $0ijk$-plane. Therefore simultaneously imposing multiple projectors
is equivalent to the presence of multiple intersecting D3--branes. 
A list of all possible orthogonal D3--brane 
intersections which preserve some
fraction of supersymmetry can be found from the M-fivebrane
intersections given in \cite{GP,GLW}. Combined with our intuition from
\cite{core} these observations suggest that solutions to the generalised 
Nahm's equations~\reef{Beq} can be interpreted as the
D1--branes opening up into a collection of intersecting D3--branes. 
That this is indeed the case will be shown in section four. However, 
before discussing explicit solutions it is instructive to consider the 
generalized Nahm's equations and the D3--brane configurations they
describe in more detail. 

In general intersecting D3--branes which preserve some fraction of the 
spacetime supersymmetry can be thought of as a single D3--brane which 
is stretched over a  three-manifold in the space in which the branes are
embedded. The condition that supersymmetry is preserved is known to be
equivalent to the statement that the manifold over which the D3--branes are 
stretched is a calibrated sub-manifold of the embedding 
space~\cite{BBS,GP,GLW,AFS}. 
Here the embedding space will simply be $\mathbb{R}^n$ for $n=3,5,6,7$
and the calibrated 
three-manifolds which we will encounter
are known as $\mathbb{R}^3$, K\"{a}hler, special Lagrangian 
and associative
submanifolds.  We recall that a calibration \cite{HL} is a closed $p$-form
$\omega$ in the bulk space with the property that, for any tangent vector
$\xi$ to a $p$-dimensional sub-manifold, $P[\omega](\xi) \le dvol(\xi)$,
where $dvol$ is the induced volume form on the sub-manifold
and $P[\omega]$ denotes the pull-back of $\omega$ to the worldvolume. 
A sub-manifold for which this inequality is saturated is said to be
calibrated. These distinguished surfaces each represent the minimal 
volume ({\it i.e.} energy) elements of their respective homology classes.

One of the main observations of this paper is that 
the constants $c_{ijk}$ (for which there are supersymmetric solutions to  
equation~\reef{Beq}) are exactly the non-vanishing components of 
the calibration forms associated
with the corresponding 3-manifold over which the D3--brane intersection is
stretched. Put differently, the generalized Nahm's equations describe 
D1--brane configurations which open up into D3--brane intersections that 
are stretched over a calibrated three-manifold whose calibration form 
$\omega$ is none other than the three-form 
$\frac{1}{3!}c_{ijk}dx^i\wedge dx^j\wedge dx^k$. Furthermore
the Bogomoln'yi bound \reef{T} can be written as 
\be
E \ge
T_1\lambda^2\int dx^9 {\rm STrP}\left[i_{\Phi}i_{\Phi}\omega\right]\ ,
\ee
where $i_{\Phi}$ is the the non-Abelian interior product introduced in 
\cite{dielec}, $STr$ is the symmeterised trace and $P$ 
denotes the pull-back to the D1--brane by the non-Abelian
scalars $\Phi^i$.
Thus the energy of the D1--brane is bounded below by 
the (non-Abelian) 
pull-back of the calibrating form to the D1--brane
worldvolume. This is in complete analogy with the
Abelian Bogomoln'yi bound on the D3--brane worldvolume \cite{BBS,GP,GLW,AFS}
and provides an new interpretation of fuzzy funnels as non-Abelian calibrated
3-surfaces. This is reminiscent of the notion of a generalised calibration
given in \cite{gencal}, which included Abelian worldvolume gauge fields.
In particular 
the S-dual configuration of a fundamental string ending on a
D$p$-brane is an example of such a (generalised) calibrated $p$-surface.

In order to demonstrate our claims in concrete examples we now summarise 
the various cases with which we will be concerned. 
In what follows the $c_{ijk}$ components with $+1$ are chosen to be so, 
whereas those equal to $-1$ are then fixed by supersymmetry. 
This change in sign corresponds
to adding an anti-D3-brane.
The simplest example is the one already discussed in
~\cite{core} for which the only non-vanishing scalars are taken to be
$\Phi^{1,2,3}$. Here the non-Abelian D1--branes open up into a single
D3-brane which spans $\mathbb{R}^3$. The relevant D3--brane configuration,
the non-vanishing components of $c_{ijk}$,
the fraction $\nu$ of preserved supersymmetries on the D1--brane
and the associated Bogomoln'yi equations \reef{Beq} can be summarised as
\ba\label{C0}
\matrix{
D3:&1&2&3&\cr
D1:&&&&&&&&9\cr}\nonumber\\
c_{123}=1\quad\quad\quad \nu=1/2\quad {}
\ea
\ba
{\Phi^1}' = [\Phi^2,\Phi^3]\ , \quad
{\Phi^2}' = [\Phi^3,\Phi^1]\ ,\quad {\Phi^3}' = [\Phi^1,\Phi^2] \ . \nonumber\\
\nonumber
\ea
Note that $c_{ijk}$ is simply $\varepsilon_{ijk}$, which is indeed the
volume form on $\mathbb{R}^3$, and the Bogomoln'yi equations 
are the standard Nahm's
equations describing BPS monopoles on $\mathbb{R}^{3}$~\cite{ded,core}. 
More interesting examples 
can be found by allowing for more scalars to be turned on. If we take
$\Phi^{1,2,3,4,5}$ to be non-trivial then there exist supersymmetric solutions
to the generalised Nahm equations describing the following configuration
\ba\label{C1}
\matrix{
D3:&1&2&3&   \cr
D3:&1& & &4&5\cr
D1:&&&&&&&&9\cr}\nonumber\\
c_{123}=c_{145}=1\quad \nu=1/4\quad{}
\ea
\ba
{\Phi^1}' &=& [\Phi^2,\Phi^3]+[\Phi^4,\Phi^5]
\ , \nonumber\\
{\Phi^2}' &=& [\Phi^3,\Phi^1]\ ,\quad {\Phi^3}' = [\Phi^1,\Phi^2]
\ , \nonumber\\
{\Phi^4}' &=& [\Phi^5,\Phi^1]\ ,\quad {\Phi^5}' = [\Phi^1,\Phi^4]\ ,
\nonumber\\
{}[\Phi^2,\Phi^4] &=& [\Phi^3,\Phi^5]\ ,\quad 
[\Phi^2,\Phi^5]\,=\,[\Phi^4,\Phi^3]\ .
\nonumber\\ 
\nonumber
\ea
Supersymmetry tells us that such an D3--brane intersection 
should be stretched over $\mathbb{R} \times \cal{M}$ where $\mathbb{R}$ is the
common direction and $\cal{M}$ is a
complex curve. 
Such complex curves are calibrated by the K\"{a}hler form associated with a 
given complex structure on the four manifold spanned by $\Phi^{2,3,4,5}$.
Forming the complex pairs $Z^1=\Phi^2+i\Phi^3$ and $Z^2=\Phi^4+i\Phi^5$ we see 
that $c_{ijk}dx^i\wedge dx^j\wedge dx^k$ 
in this case is nothing but the wedge product of $dx^1$ with 
the K\"{a}hler form associated with this complex structure.

There are two distinct ways to obtain configurations which preserve $\nu = 1/8$
of the supersymmetry. The first is a straightforward generalisation of the
K\"{a}hler case above which is obtained by turning on 
$\Phi^{1,2,3,4,5,6,7}$. The corresponding D3--brane intersection 
and generalized Nahm's equations are
\ba\label{C2}
\matrix{
D3:&1&2&3&       \cr
D3:&1& & &4&5    \cr
D3:&1& & & & &6&7\cr
D1:&&&&&&&&&9\cr}\nonumber\\
c_{123}=c_{145}=c_{167}=1\quad \nu=1/8\quad{}
\ea
\ba
{\Phi^1}' &=& [\Phi^2,\Phi^3]+[\Phi^4,\Phi^5]+[\Phi^6,\Phi^7]
\ , \nonumber\\
{\Phi^2}' &=& [\Phi^3,\Phi^1]\ ,\quad {\Phi^3}' = [\Phi^1,\Phi^2]
\ , \nonumber\\
{\Phi^4}' &=& [\Phi^5,\Phi^1]\ ,\quad {\Phi^5}' = [\Phi^1,\Phi^4]
\ , \nonumber\\
{\Phi^6}' &=& [\Phi^7,\Phi^1]\ ,\quad {\Phi^7}' = [\Phi^1,\Phi^6]
\ , \nonumber\\
{}[\Phi^2,\Phi^4] = [\Phi^3,\Phi^5]\ ,&& [\Phi^2,\Phi^5]=[\Phi^4,\Phi^3]\ ,
\,\quad \,[\Phi^2,\Phi^6]=[\Phi^3,\Phi^7]\ ,\nonumber\\
{}[\Phi^2,\Phi^7] = [\Phi^6,\Phi^3]\ ,&& [\Phi^4,\Phi^6]=[\Phi^5,\Phi^7]\ ,
\,\quad \,[\Phi^4,\Phi^7]=[\Phi^6,\Phi^5]\ .\nonumber\\
\nonumber
\ea
Once again this intersection should be stretched over 
$\mathbb{R} \times \cal{M}$ where now $\cal{M}$ is to be regarded as a complex
curve embedded into six dimensional Euclidean space.
Forming complex pairs as $Z^1=\Phi^2+i\Phi^3$, $Z^2=\Phi^4+i\Phi^5$
and $Z^3=\Phi^6+i\Phi^7$ we see that  $c_{ijk}dx^i\wedge dx^j\wedge dx^k$ 
in this case is again the wedge product of $dx^1$ with 
the K\"{a}hler form.
A more interesting example which preserves the same amount of supersymmetry
is provided by turning on $\Phi^{1,2,3,4,5,6}$ as follows
\ba\label{SLAG}
&&\matrix{
D3:&1&2&3&       \cr
D3:& &2& &4& &6  \cr
D3:& & &3& &5&6   \cr
\bar{D3}:&1& & &4&5& \cr
D1:&&&&&&&&&9\cr}\nonumber\\
c_{123}&=&c_{145}\,=\,c_{246}\,=\,-c_{356}\,=\,1 \quad \nu=1/8 
\ea
\begin{eqnarray}
{\Phi^1}' &=& [\Phi^2,\Phi^3]+[\Phi^4,\Phi^5]\ , \nonumber\\
{\Phi^2}' &=& [\Phi^3,\Phi^1]+[\Phi^4,\Phi^6]\ ,\nonumber\\
{\Phi^3}' &=& [\Phi^1,\Phi^2]+[\Phi^6,\Phi^5]\ , \nonumber\\
{\Phi^4}' &=& [\Phi^5,\Phi^1]+[\Phi^6,\Phi^2]\ ,\nonumber\\
{\Phi^5}' &=& [\Phi^1,\Phi^4]+[\Phi^6,\Phi^3]\ , \nonumber\\
{\Phi^6}' &=& [\Phi^2,\Phi^4]+[\Phi^5,\Phi^3]\ , \nonumber\\
{}[\Phi^2,\Phi^5]&+&[\Phi^3,\Phi^4]+[\Phi^6,\Phi^1]=0\ .\nonumber\\
\nonumber
\ea
Note that there is no direction shared by all of the D3--branes. Supersymmetry
implies that this intersection should be stretched over a special Lagrangian
three-manifold embedded into six dimensional Euclidean space.
Defining our complex coordinates to be $Z^1=\Phi^1+i\Phi^6$, 
$Z^2=\Phi^2+i\Phi^5$ and $Z^3=\Phi^3+i\Phi^4$ we may introduce the 
holomorphic three form $\psi = dZ^1\wedge dZ^2\wedge dZ^3$. It is then
straightforward to see that $c_{ijk}$ represents the non-vanishing components
of $\omega = Re(\psi)$ which is the calibration form for a 
special Lagrangian surface embedded into the Euclidean six-dimensional 
space endowed with the above complex structure~\cite{HL}.  

The final example again involves the seven scalars $\Phi^{1,2,3,4,5,6,7}$ 
and leads to the following configuration
\ba\label{G2}
&&\matrix{
D3:&1&2&3&     \cr
\bar{D3}:& & &3&4& & &7\cr
\bar{D3}:& & &3& &5&6\cr
D3:&1& & & & &6&7\cr
D3:&1& & &4&5& \cr
D3:& &2& &4& &6\cr
\bar {D3}:& &2& & &5& &7\cr
D1:&&&&&&&&&9\cr}\nonumber\\
c_{123}=c_{145}&=&c_{167}=c_{246}=-c_{257}=-c_{347}=-c_{356}=1
\quad \nu=1/16
\ea
\begin{eqnarray}
{\Phi^1}' &=& [\Phi^2,\Phi^3]+[\Phi^4,\Phi^5]+[\Phi^6,\Phi^7]\ , \nonumber\\
{\Phi^2}' &=& [\Phi^3,\Phi^1]+[\Phi^4,\Phi^6]+[\Phi^7,\Phi^5]\ ,\nonumber\\
{\Phi^3}' &=& [\Phi^1,\Phi^2]+[\Phi^7,\Phi^4]+[\Phi^6,\Phi^5]\ , \nonumber\\
{\Phi^4}' &=& [\Phi^5,\Phi^1]+[\Phi^6,\Phi^2]+[\Phi^3,\Phi^7]\ ,\nonumber\\
{\Phi^5}' &=& [\Phi^1,\Phi^4]+[\Phi^3,\Phi^6]+[\Phi^2,\Phi^7]\ , \nonumber\\
{\Phi^6}' &=& [\Phi^2,\Phi^4]+[\Phi^5,\Phi^3]+[\Phi^7,\Phi^1]\ , \nonumber\\
{\Phi^7}' &=& [\Phi^1,\Phi^6]+[\Phi^5,\Phi^2]+[\Phi^4,\Phi^3]\ . \nonumber\\
\nonumber
\end{eqnarray}
Here the components of the three form $c_{ijk}$ are precisely the 
non-zero components
of the unique three form in $\mathbb{R}^7$ which is invariant 
under the exceptional group $G_2$.
These particular $c_{ijk}$ can be identified with the
octonionic structure constants 
and the calibration form $\omega$ calibrates so-called associative 
three-surfaces in $\mathbb{R}^7$~\cite{HL}.

In what follows we will refer to these various cases as $\mathbb{R}^3$, 
K\"{a}hler, special Lagrangian and associative respectively. 
Before proceeding let us make several comments on the generalised Nahm's
equations. 

First 
we note that these equations are not new but can be recognised as the 
dimensional reduction of a higher-dimensional 
self-duality condition \be
F_{IJ} = \frac{1}{2}t_{IJKL}F^{KL}\ ,\label{higherD}
\ee
where $\{x^I\} = \{x^i,x^9\}$  and $t_{9ijk}=c_{ijk}$, $t_{ijkl}=0$ 
\cite{Corrigan}. In particular the generalised Nahm's equations arise by
assuming that the gauge field $A_I$ depends only
on $x^9$ and choosing the gauge $A_9=0$. These equations were
previously analysed in ~\cite{Fairlie,Ueno,Ivanova,BF,Sfetsos,BLP}.

Next we note that in the $\mathbb{R}^3$ 
and associative cases there is no constraint equation
since there are no pairs $i,j$ such that
$c_{ijk}=0$ for all $k$. Further, in these cases the $c_{ijk}$ satisfy
\be\label{antsym} 
c_{ijk}c_{lmk}=\delta_{il}\delta_{jm}-\delta_{jl}\delta_{im} +\gamma_{ijlm}
\ee
where $\gamma_{ijlm}$ is antisymmetric in $i,j,l,m$.
Hence \reef{constraint2} follows from the Jacobi identity.

As a final comment we note that in all of the above cases the D1--brane
should appear as a monopole on the calibrated surface. 
Applying S-duality to our configurations merely has the effect of
changing the D1--branes into fundamental strings. By definition
a D3--brane is a suitable end point for a fundamental string, 
and the condition that it is calibrated implies that some supersymmetry is
preserved. Therefore  we expect that smooth solutions
representing D1--branes, {\it i.e.,} monopoles,  on calibrated three-surfaces
exist. Further, T-dualising along the
$x^1,x^2,x^3$ directions produces a configuration of intersecting
D4-branes with a single D0--brane (corresponding to the first D3-brane). 
This system should therefore 
correspond to an instanton on a calibrated four-surface and are related
to non-trivial solutions to \reef{higherD}.
Indeed it was shown in \cite{AFSO} that supersymemtric 
wrapped D--branes in manifolds of
special holonomy give rise to cohomological field theories whose equations
of motion localise to solutions of \reef{higherD}.
Starting with the above equations  such
a class of solutions is obtained by taking $\partial_{x^9}\Phi^i=0$. Hence
the above equations all become constraints on the commutators.
These solution may therefore be interpreted as supersymmetric D0-brane
states and in particular we expect that the examples 
of  K\"ahler calibrations are related to the construction of
complex curves in (M)atrix theory \cite{CT}.

\sect{Solutions}

In this section we will provide some explicit solutions to the 
generalised Nahm's equations which can be interpreted as fuzzy funnels that
open up into the D3--brane intersections discussed in the previous section. 
Our focus here is to demonstrate that solutions to the generalised Nahm's 
equations do in fact exist and describe D1--branes ending on supersymmetric
D3--brane intersections. We therefore present only a set of very 
simple configurations which correspond to the D3--brane intersections at the
origins of both the Higgs and Coulomb branches. We will argue that the
generalised Nahm's equations capture all of the physics of these intersections
in the following section when we analyze deformations of the solutions 
presented here.\footnote{We note that a general 
family of solutions to the associative example were 
obtained in \cite{Fairlie,Ueno,BF,Sfetsos}. 
However the fields were not in $u(N)$ and hence they cannot be readily
embedded into the D1--brane effective action.}

For the purposes of orienting the reader we will begin by briefly 
reviewing the fuzzy funnel solutions to the basic Nahm equations appearing in 
equation~\reef{C0} which were originally  presented in \cite{core}. 
Taking only $\Phi^{1,2,3}$ to be non-vanishing we make the ansatz
\be\label{ansatz1}
\Phi^i=f(x^9)\alpha^i\ ,
\ee
where $[\alpha^i,\alpha^j]=2\varepsilon^{ijk}\alpha^{k}$ is an $n$-dimensional
representation of $su(2)$. This is easily seen to solve 
the Nahm's equations so long as $f^{\prime}=2f^2$ which gives
\be\label{sol0}
\Phi^{i}=-\frac{1}{2}\frac{1}{x^9-a}\alpha^{i}\ ,
\ee
where $a$ is an arbitrary constant of integration. The profile of this 
solution is clearly that of a fuzzy, or $su(2)$ valued, funnel 
which opens up into a three dimensional surface {\it i.e.,} $\mathbb{R}^3$, 
as $x^9\rightarrow a$. At each finite value of $x^9$ the cross section
of the funnel is a fuzzy two-sphere.  The $x^9$ dependent radius of 
the funnel is given by
\be\label{rad0}
R(x^9)^2\equiv -\frac{\lambda^2}{n}\sum_{i=1}^{3}{\rm Tr}[\Phi^i(x^9)^2] = 
\frac{c_2\lambda^2}{4}\frac{1}{(x^9-a)^2}
\ee
where $c_2$ is the quadratic Casimir of the $su(2)$ representation.
We will work with the $n$-dimensional irreducible representation 
for which we have $c_2=n^2-1$.       

The energy of this configuration may be obtained by evaluating equation~\reef{E}
on the solution presented in equation~\reef{sol0}. We find
\be\label{E0}
E=\frac{T_1\lambda^2}{3}\int dx^9 c_{ijk}
{\rm Tr}\left(\Phi^i\Phi^j\Phi^k\right)^{\prime} = 
T_3\left(1-1/n^2\right)^{-1/2}\int 4\pi R^2dR\ .
\ee
where we have used $T_1=T_3(2\pi l_s)^2$ and we have identified the physical
radius of the funnel $R$ with the radial coordinate in the space spanned
by $x^1,x^2,x^3$. It is now clear that for large $n$ the energy of the funnel 
configuration can be identified with the energy of a single, flat D3-brane 
sitting at $x^9=a$ filling the $x^1,x^2,x^3$ directions. To further support
the claim that the funnel is indeed opening up into a D3--brane
we note that the non-Abelian Wess-Zumino couplings identified in 
\cite{dielec,watimark} can be evaluated as
\ba\label{RRcharge}
i\lambda\mu_1\int {\rm STrP}[i_{\Phi}i_{\Phi}C^{(4)}] &=&
\frac{i\lambda}{2}\mu_1\int dtdx^9\,C^{(4)}_{tkji}{\rm Tr}
\left(\Phi^{k\prime} [\Phi^i,\Phi^j]\right) \nonumber\\
&=&i\mu_3\left(1-1/n^2\right)^{-1/2}\int 4\pi R^2dR \,C_{0123}
\ea
indicating that the funnel is acting as a source for precisely the correct
Ramond-Ramond four form field to be identified as a D3--brane.

We will now discuss solutions to the generalised Nahm 
equations. It will be shown below that the K\"ahler and special 
Lagrangian cases can all be obtained as special cases of the 
associative example. Thus we will begin our analysis
with the configuration in~\reef{G2}. 
The generalized Nahm equations in \reef{G2} can be solved by taking 
\be\label{G2anz}
\Phi^{i} = f(x^9)A^i\ ,
\ee 
where $A^i$ are a set of $N\times N$ constant matrices which satisfy 
$\frac{1}{4}c_{ijk}[A^j,A^k]=A^i$. A solution is given by
\ba\label{EA}
A^1&=&{\rm diag}(\alpha^1,0,0,\alpha^1,\alpha^1,0,0) 
\nonumber\\ 
A^2&=&{\rm diag}(\alpha^2,0,0,0,0,\alpha^1,\alpha^1)
\nonumber\\
A^3&=&{\rm diag}(\alpha^3,\alpha^1,\alpha^1,0,0,0,0) 
\nonumber\\
A^4&=&{\rm diag}(0,\alpha^3,0,0,\alpha^2,\alpha^2,0)
\nonumber\\
A^5&=&{\rm diag}(0,0,\alpha^3,0,\alpha^3,0,\alpha^3) 
\nonumber\\
A^6&=&{\rm diag}(0,0,\alpha^2,\alpha^2,0,\alpha^3,0) 
\nonumber\\
A^7&=&{\rm diag}(0,\alpha^2,0,\alpha^3,0,0,\alpha^2)\ ,
\ea
where $\alpha^a$ satisfy $[\alpha^a,\alpha^b]=2\varepsilon^{abc}\alpha^c$
are now regarded as an $n$ dimensional representation of $su(2)$ so that 
$N=7n$. Notice that each diagonal block contains exactly one copy 
of the $su(2)$ generators. 
Substituting this ansatz into the generalised Nahm equation 
leads to  $f'=2f^2$.  Hence we again find funnel-like solutions of the form 
\be
\Phi^i = -{1\over 2}{1\over x^9-a}A^i\ .
\label{simsol}
\ee 
Analogous block diagonal 
solutions to the higher-dimensional self-duality equation 
\reef{higherD} have previously been constructed \cite{MrB}. These have the
interpretation of four-dimensional instantons embedded onto calibrated
four-surfaces in $\mathbb{R}^8$.

It is easy to see that
\be
(\Phi^1)^2+\ldots +(\Phi^7)^2 = -{c_2\over 4}{1\over (x^9-a)^2}
{\rm diag}(1,1,1,1,1,1,1) \ ,
\label{funnel}
\ee
where $c_2 = -{1\over n}{\rm Tr}((\alpha^1)^2+(\alpha^2)^2+(\alpha^3)^2)$
is the Casimir invariant of $su(2)$ in our $n$-dimensional representation.
In this case the D1--branes should not be thought of as expanding into
a fuzzy six sphere\footnote{We would like to thank R. Myers for a 
discussion on this point.}. To see this recall that each block in 
equation~\reef{EA}
contains exactly one complete set of the $su(2)$ generators. 
As a result each block gives rise to a single copy of the 
fuzzy funnel described above. 
This configuration should therefore be viewed as expanding into 
seven intersecting fuzzy two spheres each of 
which, as $x^9\rightarrow a$, become one of the D3--branes making up the 
intersection of~\reef{G2}. 

In fact the ansatz made in equation~\reef{G2anz} can
easily be generalised to include a different profile for each diagonal block.
This leads to seven independent integration constants and therefore corresponds
to orthogonal D3--branes which are separated in the $x^9$ direction. 
One may also trivially add an identity matrix to each diagonal block. This
corresponds to separating the fuzzy funnels in directions orthogonal to the 
D1--branes. In this section we will restrict ourselves to discussing the case 
where all of the D3--branes are located at the same position 
in $x^9$ as well as in the transverse directions---see the following section for
more details.

Under these assumptions we may take
\be
R^2(x^9)=\frac{c_2\lambda^2}{4}\frac{1}{(x^9-a)^2}
\ee
to be the $x^9$ dependent radius of each of the fuzzy funnels. The  
energy of this configuration is then easily evaluated to be 
\be\label{Eexceptional}
E=\frac{T_1\lambda^2}{3}\int dx^9 c_{ijk}
{\rm Tr}\left(\Phi^i\Phi^j\Phi^k\right)^{\prime} = 
7T_3\left(1-1/n^2\right)^{-1/2}\int 4\pi R^2dR\ ,
\ee
which is precisely the energy one expects from the D3--brane intersection
in~\reef{G2}.
Finally we may verify our interpretation of the funnel 
by examining the induced RR couplings
\ba\label{WZ2}
i\lambda\mu_1\int {\rm STrP}[i_{\Phi}i_{\Phi}C^{(4)}]&=&
\frac{i\lambda}{2}\mu_1\int dtdx^9\,C^{(4)}_{tkji}{\rm Tr}
\left(\Phi^{k\prime} [\Phi^i,\Phi^j]\right) \nonumber\\
&=&i\mu_3\left(1-1/n^2\right)^{-1/2}\int 4\pi R^2dR\left( 
C_{0123}+C_{0145}+C_{0167}+C_{0246}\right.\nonumber\\
&&\ \ \ \ \ \ \ \ \ \ \ \ \ \ \ \ \ \ \ \ \ \ \ \ \ \ \ \ \ \ \ \ \ \ 
-\left.C_{0257}-C_{0347}-C_{0356}\right)\ 
\ea
which indicates that this non-Abelian embedding of D1--branes
is a source for precisely the correct Ramond-Ramond fields 
to be identified with the configuration~\reef{G2}.

In order to find similar solutions to the other generalised Nahm equations we
observe that all of the equations 
(and constraints) presented in~\reef{C1}, \reef{C2} and \reef{SLAG} 
can be viewed as special cases of the final, associative case, \reef{G2}. 
In particular the special Lagrangian 
case \reef{SLAG} follows by setting $\Phi^7=0$ and the constraint
arises from the $\Phi^7$ equation in \reef{G2}. A solution to these equations
is then found by simply deleting the second, fourth and seventh blocks 
from each of the matrices in equation~\reef{EA}. 
We find $\Phi^i=f(x^9)A^i$ where
\ba\label{SLAGA}
A^1&=&{\rm diag}(\alpha^1,0,\alpha^1,0) \nonumber\\ 
A^2&=&{\rm diag}(\alpha^2,0,0,\alpha^1) \nonumber\\
A^3&=&{\rm diag}(\alpha^3,\alpha^1,0,0) \nonumber\\
A^4&=&{\rm diag}(0,0,\alpha^2,\alpha^2) \nonumber\\
A^5&=&{\rm diag}(0,\alpha^3,\alpha^3,0) \nonumber\\
A^6&=&{\rm diag}(0,\alpha^2,0,\alpha^3)\ . 
\ea
Assuming the same profile for each complete set of $su(2)$ 
generators this solution corresponds to a collection of 
six intersecting fuzzy funnels which open into the D3--brane 
intersection described in~\reef{SLAG}. 
On the other hand by imposing the constraint relations in \reef{C2} directly on 
the equations in \reef{G2} we obtain the generalised 
Nahm equations corresponding to the K\"ahler case in \reef{C2} 
which have the simple solutions $\Phi^i=f(x^9)A^i$ with
\ba\label{CB}
A^1&=&{\rm diag}(\alpha^1,\alpha^1,\alpha^1) \nonumber\\ 
A^2&=&{\rm diag}(\alpha^2,0,0) \nonumber\\
A^3&=&{\rm diag}(\alpha^3,0,0) \nonumber\\
A^4&=&{\rm diag}(0,\alpha^2,0) \nonumber\\
A^5&=&{\rm diag}(0,\alpha^3,0) \nonumber\\
A^6&=&{\rm diag}(0,0,\alpha^2) \nonumber\\
A^7&=&{\rm diag}(0,0,\alpha^3)\ ,
\ea
representing three intersecting fuzzy funnels which this time
open into the D3--brane intersection of~\reef{C2}.

Lastly by setting $\Phi^6=\Phi^7=0$ in~\reef{C2} we obtain the equations
and constraints of the K\"ahler intersection in~\reef{C1}. 
These are solved by truncating the matrices
in equation~\reef{CB} in the obvious way to give $\Phi^i=f(x^9)A^i$ where
\ba\label{CA}
A^1&=&{\rm diag}(\alpha^1,\alpha^1) \nonumber\\ 
A^2&=&{\rm diag}(\alpha^2,0) \nonumber\\
A^3&=&{\rm diag}(\alpha^3,0) \nonumber\\
A^4&=&{\rm diag}(0,\alpha^2) \nonumber\\
A^5&=&{\rm diag}(0,\alpha^3) \ .
\ea

As a final comment on all of the solutions presented in this section
we note that as the scalars $\Phi^i$ are
becoming large and quickly varying near $x^9=a$ the 
Yang-Mills approximation in which we are working is strictly 
no longer valid. On the other hand for $x^9 \gg a$ we should be able 
to trust the solution. As discussed in detail in ~\cite{core,fun} 
this is the opposite
range of approximation that is valid on the worldvolume theory of
the D3--branes that the D1--branes are intersecting. 
Indeed the picture presented here accurately describes the 
region near the intersection. 

\section{Moduli}

In the previous section we constructed explicit supersymmetric fuzzy funnel
solitons.  However these solutions 
are somewhat trivial since the various components
of the scalar fields which describe the different D3-branes commute
with each other. Indeed these solutions simply represent several
distinct fuzzy funnels, each interpreted as a group of $n$ D1--branes
ending on a single D3--brane, embedded into a suitably large matrix. 
Therefore we would like to obtain more complicated
solutions. Unfortunately finding explicit new solutions is, in general, 
a difficult task. Here we will perform an analysis of the 
linearised modes about solutions of the form  \reef{simsol}. 
In what follows we will establish the existence
of a large number of moduli which lead to non-trivial solutions, at least
at the linearised level. 

It is convenient to rewrite the solution in the form
\be
\Phi^i(x^9) = -{1\over 2x^9}A^i_a T^a
\ee
here we have placed the D3--branes at $x^9=0$ and 
$T^a$, $a=0,1,...,N^2-1$ are the generators
of $u(N)$; $[T^a,T^b] = f^{abc}T^c$ with $T^0$ taken to be 
the generator of the  overall Abelian $u(1)$ in 
$u(N) = u(1)\oplus su(n)$. In other words we have used
the fact that we must embed our matrices into $u(N)$ to express
$A^i$ as a linear combination of the generators $T^a$. 
Note that
since $\Phi^i$ solves the Bogomoln'yi equation we can deduce that
\be
A^i_a = {1\over 4}c_{ijk}f^{abc}A^j_bA^k_c\ .
\label{id}
\ee
In addition the $f^{abc}$ satisfy the Jacobi identity
$f^{abe}f^{cde}+f^{cbe}f^{dae}+f^{dbe}f^{ace}=0$.

A linearised perturbation can also be written as 
$\delta \Phi^i = \varphi^i_aT^a$ and therefore satisfies the
equation
\be
x^9 {d\varphi^i_a \over dx^9} = -{1\over 2}c_{ijk}f^{abc}A^j_b\varphi^k_c\ .
\label{zmeq}
\ee
For most examples  
we will also have to worry about the constraint equation 
\reef{condition} but here we will only consider 
the $\mathbb{R}^3$ fuzzy funnel \reef{C0} and the associative
case \reef{G2} for which the constraint is absent.
Since $c_{ijk}, f^{abc}$ and $A^j_b$ are known constant tensors we
have a linear equation for the  perturbations $\varphi^i_a$.
Furthermore, since $(T^a)^\dag=-T^a$ and 
$c_{ijk}f^{abc}A^j_b= c_{kji}f^{cba}A^j_b$, the
matrix on the right hand side is real and symmetric. It follows that there
is a basis of solutions with eigenvalues $\lambda_a^i$ and hence
there are $DN^2$ zero-modes
\be
\varphi^i_a = \epsilon^i_a (x^9)^{\lambda^i_a}\ ,
\ee
where $\epsilon^i_a$ is a small parameter. Note that the perturbation
is only valid if the second order term in the variation (there are no
higher order terms) is small compared to the linear one. 
This corresponds to 
$\epsilon^i_a << (x^9)^{-\lambda^i_a-1}$.
So any given perturbation  
isn't valid over the entire funnel but is always valid over some
region. Strictly speaking 
we may only trust the Yang-Mills
approximation when the derivatives of the gauge fields are small,
corresponding to the region $x^9\rightarrow \infty$.
Therefore 
we should only trust zero-modes with $\lambda^i_a \le -1$, although we
will see exceptions to this.
We would also like to see if some of the moduli are associated with
the D1--brane itself and if some in fact reflect moduli of the intersecting
D3-branes. We expect that zero-modes which do not vanish as
$x^9\rightarrow \infty$ correspond to moduli of the D1--branes whereas
those that do vanish correspond to the geometry of the D3--branes.

Let us start by identifying some obvious zero-modes.
Firstly we can separate out $D$ center of mass coordinates along each 
of the transverse directions $x^i$. This corresponds simply to taking 
$\varphi^i$ to be constant and proportional to $T^0$ 
({\it i.e.} $\lambda = 0$).
Next we can identify the translation mode 
$\varphi^i_a = \epsilon {\Phi^i}_a' = {1\over 2}\epsilon A^i_a/(x^9)^2$ 
({\it i.e.} $\lambda  = -2$). 
That this solves \reef{zmeq} follows from the identity \reef{id}.
This mode therefore represents the location of the D3--brane along
the $x^9$ direction. 
Finally it is clear that we can act by gauge transformations on the
original solution; $\Phi^i\rightarrow g\Phi^i g^{-1}$, $g\in SU(N)$.
Expanding $g=e^{h_a T^a}$ with $h_a<<1$  
leads to $N^2-1$ zero-modes $\varphi^i_a = -{1\over 2}f^{abc}h_bA^i_c/x^9$,
({\it i.e.} $\lambda  = -1$).
Using the Jacobi identity and \reef{id} one sees
that this indeed solves \reef{zmeq}.

\subsection{$\mathbb{R}^3$}

Before analysing the associative case 
it is instructive  to analyse the simplest case  of the original
fuzzy funnel with $D=3,N=2$, $c_{ijk}=\varepsilon_{ijk}$ and 
$f^{abc}=2\varepsilon^{abc}$ if $a,b,c\ne 0$, $f^{0bc}=0$. Thus 
there are   12 zero-modes. The solution is given in 
\reef{ansatz1} and has  $A^i=\alpha^i$ so that 
$A^i_a = \delta^i_a$. As mentioned above we can identify
3 zero-modes $\varphi^i = c^iT^0$ 
corresponding to the constant center of mass
coordinates. The rest of the
the zero-modes, namely those that are in $su(2)$, are represented
by square $3\times 3$ matrices $\varphi^i_j$ which satisfy
\be
x^9 {d \varphi^i_j\over d x^9} = 
-(\delta^j_i\delta^l_k-\delta^j_k\delta^l_i)\varphi^k_l\ .
\ee
We can split up $\varphi^i_j$ into its trace, antisymmetric
and symmetric-traceless parts. The trace corresponds to
$\varphi^i_j \sim \delta^i_j$ and is just the translational
zero mode given above. The antisymmetric part can be written
as $\varphi^i_j \sim \epsilon^{ijk}h_k$ and can be seen to 
correspond to the gauge transformations. 

The symmetric part is a little more interesting.
It gives five zero-modes. There are diagonal choices of the form
\be\label{newmodes}
\varphi^i_j \sim 
\left(\matrix{-1&0&0\cr0&-1&0\cr0&0&2}\right)\ ,
\left(\matrix{-1&0&0\cr0&2&0\cr0&0&-1}\right)\ ,
\left(\matrix{2&0&0\cr0&-1&0\cr0&0&-1}\right)\ ,
\ee
note that although 
there are three such modes only two are linearly independent.
The corresponding eigenvalue is $\lambda=1$ 
so that $\varphi \sim x^9$. In fact we can find the exact solution
corresponding to this deformation. In particular consider 
solutions  of the form $\Phi^i = f_i(x^9)\alpha^i$.
The Bogomoln'yi equation gives
\be
{f_1}'=2f_2 f_3\ ,\quad {f_2}'=2f_3 f_1\ ,\quad 
{f_3}'=2f_1 f_2\ .
\ee
A one parameter solution to these equations is \cite{BPP}
\be
f_1 = -\frac{d}{2}{{\rm cn}(dx^9) \over {\rm sn}(dx^9)}\ ,\quad
f_2 = -\frac{d}{2}{{\rm dn}(dx^9) \over {\rm sn}(dx^9)}\ ,\quad
f_3 = -\frac{d}{2}{1\over {\rm sn}(dx^9)}\ ,
\label{monosol}
\ee
where ${\rm sn},{\rm cn}$ and ${\rm dn}$ are Jabocbi's elliptic
functions with parameter $k$.  
Expanding in powers of $d$ one finds that 
\reef{monosol} corresponds to a linear combination of the 
zero-modes in \reef{newmodes}.
Clearly one other  linearly independent solution 
can be found in an analogous manner. These functions are generically
periodic in $x^9$ and hence have other poles, indeed they
arise in the Nahm construction of charge-two monopoles \cite{BPP}. 
Since we are  primarily
interested in solutions which are well-behaved as $x^9\rightarrow\infty$ 
we must restrict to the case $k=1$ where the period diverges and
the solution becomes
\begin{equation}
f_1=f_2 = -{d\over 2}{1\over {\rm sinh}(dx^9)}\ ,\quad 
f_3 = -{d\over 2}{{\rm cosh}(dx^9)\over {\rm sinh}(dx^9)}\ .
\label{sepmodtwo}
\end{equation}
Thus as $x^9\rightarrow\infty$, 
$\Phi^3$ tends to the constant matrix $-\frac{d}{2}\alpha^3$, while 
$\Phi^1,\Phi^2\rightarrow 0$. 
Clearly the three modes in \reef{newmodes} are interpreted as separating 
the D1--branes along the $x^i$ axis at infinity.

This leaves three zero-modes of the form
\be\label{odd}
\varphi^i_j \sim 
\left(\matrix{0&1&0\cr1&0&0\cr0&0&0}\right)\ ,
\left(\matrix{0&0&1\cr0&0&0\cr1&0&0}\right)\ ,
\left(\matrix{0&0&0\cr0&0&1\cr0&1&0}\right)\ .
\ee
These also correspond to eigenvalue $\lambda=1$. Let us again look
for the corresponding exact solutions of Nahm's equation. In particular
we consider the first zero-mode in \reef{odd} and let
\ba
\Phi^1 = g_1(x^9)\alpha^1 + g_2(x^9)\alpha^2\ ,\quad
\Phi^2 = g_1(x^9)\alpha^2 + g_2(x^9)\alpha^1\ ,\quad
\Phi^3 = g_3(x^9)\alpha^3\ .
\label{try}
\ea 
This leads to the equations $g_1'=2g_1g_3\ ,\  g_2'=-2g_2g_3$ and 
$g_3'=2(g_1^2-g_2^2)$.
The exact solution is provided by the one parameter family
\ba
g_1 = -{b\over 4}{{\rm dn}(bx^9)+ {\rm cn}(bx^9)\over 
{\rm sn}(bx^9)}\ ,\quad
g_2= -{b(1-k^2)\over 4}{{\rm sn}(bx^9)\over 
{\rm dn}(bx^9)+ {\rm cn}(bx^9)}\ ,\quad
g_3 = -\frac{b}{2}{1
\over {\rm sn}(bx^9) }\ .
\label{exactsol}
\ea
If we expand \reef{exactsol} around $b=0$ with $k\ne 1$ then we reproduce
the first zero-mode in \reef{odd} but 
mixed with the first zero-mode in  \reef{newmodes}. Nevertheless these
solutions, along with 
similar solutions where the ansatz for $\Phi^1,\Phi^2$ and
$\Phi^3$ are permuted, 
provide three more linearly independent zero-modes. 
Again these solutions are generically periodic in $x^9$. When 
$k= 1$ the period is infinite
and we in fact find the solution \reef{sepmodtwo} again.

In summary we have found 12 zero modes. Three of these correspond to gauge
transformations and can be discarded. One of these corresponds to translations
of the D3--brane in the $x^9$ direction. This leaves $8$ linearly independent
zero modes. To understand these we can
consider the case of two D1--branes suspended between two
parallel D3--branes separated by a distance $v$. 
In this case we must take  $\Phi^i$ to have  poles at $x^9=0$ and
$x^9=v$.  These boundary conditions are 
an important part of the Nahm constuction and
was derived from string theory in \cite{KS,T}. This coincides
with Nahm's construction of charge two $SU(2)$ monopoles on 
$\mathbb{R}^3$ \cite{nahm} 
(for a helpful and more recent discussion see \cite{Sutcliffe}).
Our counting of $8$ physical zero-modes then 
agrees with the dimension of the moduli space of charge two $SU(2)$
monopoles. In particular one can see following the discussion in 
\cite{Sutcliffe} that  \reef{monosol} and \reef{exactsol} 
lead to monopoles which are separated along
the $x^1$ and $x^1=x^2 ,\ x^3=0$ axis' respectively.

\subsection{Associative}

Next we consider the associative example with $D=7,N=14$. Thus there are some
$1372$ zero-modes. The large number of zero-modes is related to the
large $N$ that we have used to embed our solution and
most zero-modes don't reflect any physics but rather the possible
embeddings. To proceed it is helpful to split the moduli space into the form
\begin{equation}
{\cal M} = {\cal M}_G\times {{\cal M}}_{\mathbb{R}^3} \times 
\tilde {\cal M} \ ,
\end{equation}
where ${\cal M}_G$ consists of the $14^2-1=195$ 
zero-modes corresponding to gauge transformations and do not
represent any physical degrees of freedom. 

The next part of the moduli space, ${\cal M}_{\mathbb{R}^3}$, is obtained by
embedding the zero-modes of a single $\mathbb{R}^3$ fuzzy funnel
into $u(14)$.
The $9$ $su(2)$ zero-modes $\varphi^i_j$ we found in section 5.1  
can  be embedded 
separately into each of the seven block diagonal entries of \reef{EA}.
However $3$ of these are gauge transformations and have already
been included ${\cal M}_G$. 
Thus the $su(2)$ zero modes give an additional $42$ moduli. 
In addition we  may also embed the $u(1)$ center
of mass zero-modes that we discussed above into $u(14)$. 
In particular if we let 
\be\label{ghgh}
\varphi^i \sim i{\rm diag}(c^i_1,c^i_2,...,c^i_7) \ ,
\ee 
where $c^i_j$ are real,
then clearly $[\varphi^i,A^j]=0$ and so \reef{zmeq} is solved if 
$\varphi^i$ is a constant, 
{\it i.e.} $\lambda=0$. This gives another $49$ zero-modes which represent
the center of mass of the D1--branes along each of the directions
and each of the D3--branes. 
In total there are $91$ linearly independent 
moduli in ${\cal M}_{\mathbb{R}^3}$.
Clearly none of these zero-modes lead to interesting new solutions. 
Instead they affect one of
the component fuzzy funnels  but leave the others invariant. 
In particular they don't lead to a mixing or
interaction between the various off diagonal blocks. 
These moduli simply represent the relative positions of the D1--branes and 
D3--branes.

We are however still left with $1086$ zero-modes and these form the
rest of the moduli space $\tilde {\cal M}$.
To count the number of physically distinct moduli we 
note that if $g$ has the
form $g = {\rm diag}(e^{i\theta_1},...,e^{i\theta_7})$,  
$\theta_1+...+\theta_7=0$ then 
$gA^ig^{-1}=A^i$. Thus if $\varphi$ is a zero-mode
then so is $g\varphi g^{-1}$ (with the same value of $\lambda$). 
Hence $\tilde {\cal M}$ has a residual gauge symmetry $U(1)^6$ 
and the physical part of the moduli space
in fact has a quotient form ${\cal N}=\tilde{\cal M}/U(1)^6$.
Note that ${\cal M}_G$ is mapped to itself under $U(1)^6$ and 
${\cal M}_{\mathbb{R}^3}$  is invariant.
However we do not know what the  fixed points of $U(1)^6$ in
$\tilde {\cal M}$ are, for example acting with $U(1)^6$
need not always produce as many as $6$ new linearly independent zero-modes.
Therefore we are unable to determine the dimension of ${\cal N}$,
although it must be {\it at least} as big as $1086/7\approx 156$.
These zero-modes cannot
be interpreted as moduli that affect only one of the component fuzzy funnels,
{\it i.e.} the corresponding solutions  are not 
simply $7$ independent fuzzy funnels but rather involve off diagonal blocks
and  represent deformations which smooth out the D3--brane intersection.

We should now compare this with the number of zero modes one expects from
such a D1/D3--brane intersection.
From the intersecting D3--branes
there are $4\cdot 21=84$ scalar modes from the hypermultiplets
arising from D3--D3 strings with $4$ DN directions. There are also
$7\cdot 6=42$ zero-modes from transverse scalar fields of the
D3--branes. 
However $7$ of these moduli are simply the translational 
zero-modes of each D3--brane along $x^9$ and have already been  
included in  ${\cal M}_{\mathbb{R}^3}$. 
Another $7$ represent the locations of the
branes along $x^8$ and these have been frozen out from the whole system.
In addition any pair of D3--branes that preserve $1/8$ 
of the spacetime supersymmetry may be rotated
by a two parameter family of angles in $SU(3)\subset SO(6)$. 
These contribute $2\cdot 21=42$ zero-modes and hence the  total number of
D3--brane intersection moduli is
$154$. These  should be viewed as the moduli of the calibrated surface on
which the D3--branes are wrapped. 
Moreover,  in addition to the surface moduli we expect that there will be
relative $U(1)$ phases of the $7$ distinct component monopoles, as well
as possible Wilson line moduli.  

Unfortunately a more 
complete analysis of the $\cal N$ moduli space 
is beyond the scope of this paper. For example we have not shown
that they all lift to full solutions. It is also important to 
understand the correct boundary conditions which are crucial in the
standard Nahm construction.
However we hope to have convinced the reader that the generalised
Nahm equations do admit non-trivial solutions whose zero-modes 
are in a one-to-one correspondence with  the moduli space 
of D1--branes ending on intersecting D3--branes. 
Recall that from the spacetime point of view these configurations appear as
monopoles on calibrated 3-surfaces. Hence the moduli space of solutions
to the generalised Nahm equations is naturally identified with the
moduli space of monopoles on calibrated 3-surfaces, including the moduli
of the surface itself.

\sect{Summary and Comments}

In this paper we have analysed the general condition for static supersymmetric
and non-Abelian embeddings of D1--branes into spacetime. The Bogomoln'yi
equations have the form of generalised Nahm equations and 
the resulting embedding has the interpretation  as a
non-Abelian calibrated surface. On the worldvolume theory the solutions 
correspond to  D1--branes which open up into 
non-commutative spheres ({\it i.e.}  fuzzy funnels). 
From the spacetime
point of view the D1--branes end on  configurations of intersecting
D3--branes. Although the exact fuzzy funnel solutions
we presented in section four were block diagonal, 
and so merely represent several
distinct fuzzy funnels, in section five
we  established the existence of a  large moduli space of non-trivial 
solutions at the linearised level.

In the case of the original fuzzy funnel the appearance of the Nahm
equation is not surprising. Soon after the discovery of D-branes it
was realised that a 
D1--brane suspended between two parallel D3--branes appears as a monopole
on the D3--brane worldvolume. Furthermore, introducing a 
probe brane provides an explicit realization of the Nahm construction
of monopoles  \cite{ded}. Here we have primarily examined solutions which
live on a half-line. In the Nahm construction these correspond
to unphysical $U(1)$ monopoles with infinite mass. From the D-brane
perspective this infinite mass is simply due to the infinite length of
the D1--brane.  However we also discussed the 
case of parallel D3--branes, or parallel sets of intersecting D3--branes, 
and obtained solutions which correspond to finite BPS monopoles.

Therefore we expect that there is an associated role for these generalised
Nahm equations here.  The natural interpretation of such a 
system is that it encodes the data for BPS monopoles 
on the calibrated three-surface. In addition we have argued 
that the geometrical
data of the calibrated surface should also be so encoded. 
Unfortunately it is not clear to us what the
recipe is for constructing the monopole fields, or D3--brane geometry, 
from solutions to the
generalised Nahm equations. In \cite{ded} the configuration \reef{C0}
was T-dualised into D5-branes ending on a D7-brane. A D1--brane probe analysis
then shows how the monopole fields can be reconstructed from the solution and
one recovers  
the  Nahm construction \cite{nahm}. However in our case we are limited
by the fact that the equivalent analysis requires more than
ten dimensions where there is no appropriate supersymmetric Yang-Mills theory.

Finally, the configurations that we have obtained in this paper are solutions
to the super-Yang-Mills approximation to the full non-Abelian Born-Infeld
action. As our solutions are supersymmetric we expect that they
will lift to solutions of the full theory although they may receive
corrections from higher orders in the non-Abelian field strength. Indeed,
the self-duality equations~\reef{higherD}, from which the generalised Nahm
equations can be obtained by dimensional reduction, 
are known~\cite{sevrin} to be corrected at higher
orders in $\lambda=2\pi l_s^2$. It would be very interesting to determine 
whether the fuzzy funnels presented here also 
receive such corrections.

\bigskip
{\noindent \bf Acknowledgements}\\

We would like to thank B. Acharya, S. Cerkis, D-E. Diaconescu, D. Fairlie, 
C. Hofman, 
C. Houghton, P. Sutcliffe, W. Taylor, D. Tong and J. Troost for 
useful discussions.
We would especially like to thank R. Myers for his initial collaboration 
on this project. The research of NRC is supported by the NSF under grant 
PHY 00-96515, the DOE under grant DF-FC02-94ER40818 and NSERC of Canada.
NDL would like to thank the Aspen Institute for Physics, 
McGill University and the Perimeter Institute for their hospitality during
the course of this work and was partially supported by a PPARC advanced
fellowship at King's College London.

\end{document}